\documentclass[sn-mathphys,Numbered, iicol]{sn-jnl}

\usepackage{graphicx}%
\usepackage{multirow}%
\usepackage{amsmath,amssymb,amsfonts}%
\usepackage{amsthm}%
\usepackage{mathrsfs}%
\usepackage[title]{appendix}%
\usepackage{xcolor}%
\usepackage{textcomp}%
\usepackage{manyfoot}%
\usepackage{booktabs}%
\usepackage{algorithm}%
\usepackage{algorithmicx}%
\usepackage{algpseudocode}%
\usepackage{listings}%
\usepackage{subcaption}
\usepackage{multicol}
\usepackage{xcolor}%

\begin{document}

\title[Article Title]{A Comparison of Adversarial Learning Techniques for Malware Detection}


\author[1]{\fnm{Pavla} \sur{Louth\'{a}nov\'{a}}}\email{louthpav@fit.cvut.cz}

\author*[1]{\fnm{Matou\v{s}} \sur{Koz\'{a}k}}\email{matous.kozak@fit.cvut.cz}

\author[1]{\fnm{Martin} \sur{Jure\v{c}ek}}\email{martin.jurecek@fit.cvut.cz}

\author[2]{\fnm{Mark} \sur{Stamp}}\email{mark.stamp@sjsu.edu}

%\street{Th\'{a}kurova 9} \street{Street}, \postcode{10587}, 
\affil*[1]{\orgdiv{Faculty of Information Technology}, \orgname{Czech Technical University in Prague}, \orgaddress{\city{Prague}, \country{Czechia}}}

\affil[2]{\orgdiv{Department of Computer Science}, \orgname{San Jose State University}, \orgaddress{\city{San Jose}, \state{California}, \country{USA}}}


\abstract{Machine learning has proven to be a useful tool for automated malware detection, but machine learning models have also been shown to be vulnerable to adversarial attacks. This article addresses the problem of generating adversarial malware samples, specifically malicious Windows Portable Executable files. We summarize and compare work that has focused on adversarial machine learning for malware detection. We use gradient-based, evolutionary algorithm-based, and reinforcement-based methods to generate adversarial samples, and then test the generated samples against selected antivirus products. We compare the selected methods in terms of accuracy and practical applicability. The results show that applying optimized modifications to previously detected malware can lead to incorrect classification of the file as benign. It is also known that generated malware samples can be successfully used against detection models other than those used to generate them and that using combinations of generators can create new samples that evade detection. Experiments show that the Gym-malware generator, which uses a reinforcement learning approach, has the greatest practical potential. This generator achieved an average sample generation time of 5.73 seconds and the highest average evasion rate of 44.11\%. Using the Gym-malware generator in combination with itself improved the evasion rate to 58.35\%.}




\maketitle

\section{Introduction}

With the rapid development of information technology, computer systems have become increasingly important in the daily lives of people. Unfortunately, the rapid development of these technologies is accompanied by a similarly rapid increase in cyberattacks.

Malicious software (malware) is one of the most significant security threats today, comprising several different categories of malicious code, such as viruses, trojans, worms, spyware, and ransomware. To protect computers and the Internet from malware, early detection is necessary. However, this is problematic, as a large amount of new malicious code is generated every day \cite{avtest2022statistics}. Since it is not possible to analyze each sample individually, automatic mechanisms are required to detect malware.

Antivirus companies often rely mainly on signature-based detection techniques \cite{al2019review} for malware detection. Signatures are specific patterns that allow for the recognition of malicious files. For example, they can be a byte sequence, a file hash, or a string. When inspecting a file, the antivirus system compares its content with the signatures of already-known malware stored in the database. If a match is found, the file is reported as malware. Signature-based detection methods are fast and effective in detecting known malware. However, malware authors can modify their code to change the signature of the program, thereby avoiding detection. Some malware can hide in the system using various obfuscation techniques \cite{singh2018challenge}, such as encryption, oligomorphic, polymorphic, metamorphic, stealth, and packing methods, to make the detection process more difficult.

Machine learning (ML) models are commonly used today in various fields. Their application can be found, for example, in technologies such as self-driving cars, weather forecasting, face recognition, or language translation systems. Machine learning has also proved to be a useful tool for automatic malware detection \cite{singh2021survey}. Unlike the signature-based method, it is capable of detecting previously unknown or obfuscated malware. However, it can be difficult to explain why the model classifies a certain file as malicious or benign \cite{dolejvs2022interpretability}, which can cause hidden vulnerabilities that attackers can exploit.

Machine learning models are vulnerable to adversarial attacks \cite{rosenberg2021adversarial}. Attackers purposely design adversarial examples, which are deliberately designed inputs to a machine learning model, to cause the model to make a mistake in its predictions. Adversarial machine learning is a field that deals with attacks on machine learning algorithms and defenses against such attacks. 

Malware detection is thus a battle between defenders and malware authors, in which each side attempts to devise new and effective ways to outwit the other. Each detection method has its own advantages and disadvantages. In various scenarios, one method may be more successful than another. Thus, the creation of an effective malware detection method is a very challenging task, and new research and methods are necessary.

The main contributions of this paper are to compare works that focus on adversarial machine learning in the area of malware detection. Specifically,
\begin{itemize}
	\item we applied some existing methods in the field of adversarial learning to selected malware detection systems. 
	\item we combined these methods to create more sophisticated adversarial generators capable of bypassing top-tier AV products. 
	\item we evaluated the single and combined generators in terms of accuracy and usability in practice.
\end{itemize} 

The rest of the paper is organized as follows: In Section \ref{sec2}, we describe state-of-the-art techniques used to generate adversarial examples present. Section \ref{sec3} provides an overview of the publications focused on creating adversarial portable executable malware samples. Section \ref{sec:experiments} describes the experiments performed and the metrics used for evaluation. Section \ref{sec5} presents and discusses the experimental results, and Section \ref{sec6} concludes this work.

%--- BACKGROUND ---------------------------------------------------
\section{Background} \label{sec2}

In this section, we describe the different methods used to create adversarial examples. We also introduce and describe selected attacks for experimentation.

\subsection{Methods for Creating Adversarial Examples}

In this section, we describe various methods to create adversarial examples.

\subsubsection{Gradient-based Approaches}

Gradient-based methods are a popular approach to generate adversarial examples. These methods work by computing the gradient of a loss function with respect to the input data. This gradient is then used to iteratively modify the input to minimize the loss. The Fast Gradient Sign Method \cite{goodfellow2015explaining} and the Jacobian-based Saliency Map Approach \cite{papernot2016limitations} are two popular gradient-based methods used for malware generation.

Given a trained model $f$ and an input example $x$, gradient-based methods generate an adversarial example $x'$ by adding a small perturbation $\delta$ to the input that maximizes the loss function $L(f(x'), y)$, where $y$ is the true input label.

The perturbation $\delta$ is calculated as follows:

\begin{align*}
\delta = \epsilon \cdot \textrm{sign} (\nabla_x L(f(x), y))
\end{align*}
where $\epsilon$ is a small constant that controls the size of the perturbation, and the sign function is used to ensure that the perturbation has the same sign as the gradient, allowing efficient computation and ensuring that the perturbation always increases the loss.

The gradient of the loss function with respect to the input $(\nabla_x L(f(x), y))$ is computed by backpropagation through the model $f$. This gradient gives the direction in which the loss function increases the most for a small change in the input and is used to determine the direction of the perturbation.

Gradient-based attacks are performed using the addition or insertion method for perturbations generated using the gradient of the cost function. When using the append method, the data (payload) is appended at the end of the file. When the insertion method is used, the payload is inserted into the slack region where the physical size is greater than the virtual size.

\subsubsection{Generative Adversarial Network-based Approaches}

Generative adversarial networks (GANs) were developed and presented by Goodfellow et al. \cite{goodfellow2014gan} in 2014. 

GAN is a system consisting of two neural networks, a generator and a discriminator, which compete against each other. The goal of the generator is to create examples that are indistinguishable from the real examples in the training set, thus fooling the discriminator. In contrast, the objective of the discriminator is to distinguish the false examples produced by the generator from the real examples that come from the training data set, thus preventing it from being fooled by the generator. The generator learns from the feedback it receives from the discriminator's classification \cite{dutta2020generative}. 

These two neural networks are trained simultaneously. The generator is constantly improving its ability to generate realistic samples, so the discriminator must continually improve its ability to distinguish between real and generated samples. This mutual competition forces both networks to continuously improve through the training process. Once this training process is completed, the generator can be used to generate new samples that are indistinguishable from the real samples \cite{wang2017generative}.

Denote the generator as $G$ and the discriminator as $D$. As described in \cite{goodfellow2014gan}, networks $G$ and $D$ play the following two-player minimax game with value function $V(G, D)$:
\begin{align*}
\underset{G}{\text{min }} \underset{D}{\text{max }} V(D,G)=\mathbb{E}_{x\sim p_{data}(x)}[log D(x)] \\
+\mathbb{E}_{z\sim p_z (z)}[log (1-D(G(z)))]
\end{align*}
that $G$ tries to minimize, while $D$ tries to maximize. $D(x)$ is the discriminator's estimate of the probability that the original data $x$ is real, $G(z)$ is the generator's output when it receives noise $z$ as input, $D(G(z))$ is the discriminator's estimate of the probability that a synthetic sample $G(z)$ of data is real, $\mathbb{E}_x$ is the expected value over all real data instances, and $\mathbb{E}_z$ is the expected value over all generated fake instances.

\subsubsection{Reinforcement Learning-based Approaches}

Reinforcement learning (RL) is a type of machine learning technique along with supervised and unsupervised machine learning. In supervised machine learning, the model is trained using a training set of labeled examples. Based on the given inputs and expected outputs, the model creates a mapping equation that the model can use to predict the labels of the inputs in the future. In unsupervised machine learning, the model is trained only on inputs without labels. The model divides the input data into classes that have similar properties, and during prediction, the inputs are labeled based on the similarity of their properties to one of the classes. Unlike supervised and unsupervised machine learning, reinforcement learning algorithms learn by interacting with an environment and getting feedback in the form of rewards or penalties rather than relying on pre-labeled instances.

A reinforcement learning model consists of two main parts: an agent and an environment. The agent learns to perform a task through repeated interactions with the environment through trial and error. In addition to the agent and the environment, the reinforcement learning model has four main subelements: a policy, a reward signal, a value function and, optionally, an environment model.
The policy defines the behavior of the agent at a particular time. In other words, it is a strategy that the agent uses to determine the next action based on the current state to achieve the highest reward. The reward signal is the feedback from the environment to the agent, indicating the success or failure of the agent's action in a given state. At each time step, the agent is in some state and sends the selected action as its output to the environment, which then returns a new state to the agent along with a reward signal.
While the reward signal shows what is beneficial in the present, the value function describes what is beneficial in the long term. The value function provides an estimate of the expected cumulative reward from the current state of the environment in the future. The agent's objective is to maximize the total reward.
The environment model mimics the behavior of the environment, making it possible to predict future states and rewards. This is an optional part of the system \cite{sutton2018reinforcement}.

Reinforcement learning is defined as repeated interactions between an agent and an environment. Individual interactions (signals exchange) are performed in time steps. At time step $t$, the environment is in state $s_t \in S$, where $S$ is the set of all possible states in the environment. The agent receives state $s_t$ and then chooses action $a_t \in A$ based on policy $\pi$, where $A$ is the set of all possible actions defined by the environment. After the environment receives information about the chosen action $a_t$ from the agent, it calculates the reward $r_t = R(s_t, a_t, s_{t+1})$ and sends it to the agent in the form of feedback. At the same time, the environment transitions to a new state $s_{t+1}$. When this cycle is complete, we say that one time step has passed. This cycle can repeat forever or end when it reaches a terminal state or a maximum time step $t = T$. We call the triplet of signals $(s_t, a_t, r_t)$ an experience. The time elapsed between $t = 0$ and the end of the environment is called an episode. A trajectory is a sequence of experiences during an episode, $\tau = (s_0, a_0, r_0),(s_1, a_1, r_1),\dots$ \cite{sutton2018reinforcement}.

In the field of creating adversarial malware samples, interactions between the agent and the environment occur in discrete time steps. The agent has a set of operations available for modifying PE files while maintaining the functionality of the malware. The goal of the agent is to perform a sequence of operations on the malware to prevent its detection.

\subsubsection{Evolutionary Algorithm-based Approaches}

Evolutionary algorithms (EA) are a useful tool for solving optimization problems. They are based on the Darwinian principle of evolution and attempt to mimic these processes. The search for the best or at least satisfactory solution to a problem takes the form of competition between gradually developing solutions within the population. Variants of EA include, for example, evolutionary strategies, evolutionary programming, genetic algorithms, and genetic programming. All of these variants share the same principle of operation but differ in their implementation.

When solving a problem, it is necessary first to define the representation of candidate solutions. These candidate solutions are called individuals or phenotypes. Since the phenotype can have a complex structure, an encoding is used to represent the individuals in an appropriate way, which is called a chromosome or genotype. Next, an initial population of individuals is created, with each individual representing a coded solution. Then, each member of the population is evaluated using a fitness function that numerically expresses the quality of the solution. The individuals with the best score are then selected and used to create a new generation. This is done using the crossover operator, which usually takes pairs of chromosomes and exchanges information between them to create new offspring. This is followed by the mutation operator, which changes a small portion of the offspring so that it is no longer just a mixture of parental genes. This introduces new genetic material into the new generation. This entire cycle (fitness evaluation, selection, crossover, mutation) is repeated until the termination condition is reached.

\subsection{Selected Attacks}

To generate samples of malicious software, we utilized three distinct techniques: gradient-based techniques, evolutionary algorithm-based techniques, and reinforcement learning-based techniques. Specifically, we selected Partial DOS \cite{demetrio2019explaining} and Full DOS \cite{demetrio2020adversarial} attacks from gradient-based techniques. From the evolutionary algorithms-based techniques, we chose GAMMA padding \cite{demetrio2021gamma} and GAMMA section-injection \cite{demetrio2021gamma} attacks. Finally, for reinforcement learning techniques, we selected the Gym-malware \cite{anderson2018gym-malware} attack. In this section, we describe each of these attacks in detail.

Partial DOS attack and Full DOS attack focus on modifying bytes in the DOS header of a portable executable (PE) file. The DOS header contains only two important pieces of information. The initial 2 bytes represent the magic number, while the final 4 bytes at offset 0x3C show the location of the PE signature in the NT headers. Partial DOS attacks modify only bytes within the range of 0x02 to 0x3B, inclusive. Full DOS attacks expand this range to include all bytes up to the PE signature. The position of this signature may vary in individual files, but it can be found at offset 0x3C.

GAMMA padding attack and GAMMA section-injection attack are based on inserting parts extracted from benign files into malware files. These are black-box attacks. Gamma attacks are formalized as a constrained optimization problem. The goal is to minimize the probability of detection but also to minimize the size of the injected content. This optimization problem is solved using a genetic optimizer. First, a random matrix is created that represents the initial population of manipulation vectors. The algorithm then iterates in three steps: selection, crossover, and mutation. During selection, the objective function is evaluated and the $N$ best candidate manipulation vectors from the current population and the population created in the previous iteration are selected. This is followed by the crossover function, where the candidates from the previous step are modified by mixing the values of pairs of randomly selected candidate vectors and a new set of $N$ candidates is returned. The last operation is a mutation, which randomly changes the elements of each vector with a low probability. At each iteration, $N$ queries are made to the target model to evaluate the objective function on new candidates and keep the best candidate population. When the maximum number of queries is reached or no further improvement in the objective function value is observed, the best manipulation vector from the current population is returned. The resulting adversarial malware sample is obtained by applying the optimal manipulation vector to the input malware sample through the manipulation operator.

The Gym-malware attack is based on reinforcement learning. The environment consists of a sample of malware and the attack target, which is an anti-malware engine. At each step, the agent receives feedback that is composed of a reward value and a vector of features that summarize the state of the environment. Based on the feedback, the agent selects mutations from a set of actions, such as adding a function to the import address table that will never be used; manipulating existing section names; creating new sections that will not be used; adding bytes to extra space at the end of sections; creating a new entry point that immediately jumps to the original entry point; manipulating debug information; packing or unpacking the file; modifying the header checksum; adding bytes to the end of the PE file. This process is repeated in several rounds, and rounds can be prematurely terminated if the agent bypasses the anti-malware engine before 10 mutations are completed.

%--- RELATED WORK -------------------------------------------------
\section{Related Work} \label{sec3}
In this section, the publications on modern methods to create adversarial examples are summarized. The section is divided into several parts, depending on the area to which the method belongs, with all publications compared based on the attacker's knowledge and strategy in Table \ref{tab:summary}.

\begin{table*}
\vspace{-0.2cm}
\caption{Summary of State-of-the-Art Adversarial Attacks against PE Malware Detection.}
\label{tab:summary}
\centering
\begin{tabular}{|c|l|l|l|l|}
  \hline
  Paper & Year & Attack framework & Knowledge & Attack Strategy \\
  \hline \hline
  \cite{castro2019aimed} & 2019 & AIMED & black-box & evolutionary algorithm\\
  \hline
  \cite{wang2020mdea} & 2020 & MDEA & black-box & evolutionary algorithm\\
  \hline
  \cite{demetrio2021gamma} & 2021 & GAMMA & black-box & evolutionary algorithm\\
  \hline
  \cite{anderson2018gym-malware} & 2018 & Gym-malware & black-box & reinforcement learning\\
  \hline
  \cite{fang2019dqeaf} & 2019 & DQEAF & black-box & reinforcement learning\\
  \hline
  \cite{song2020mab-malware} & 2020 & MAB-malware & black-box & reinforcement learning\\
  \hline
  \cite{labaca2021aimed-rl} & 2021 & AIMED-RL & black-box & reinforcement learning\\
  \hline
  \cite{kolosnjaji2018amb} & 2018 & AMB & white-box & gradient\\
  \hline
  \cite{kreuk2018deceiving} & 2018 & -- & white-box & gradient\\
  \hline
  \cite{demetrio2019explaining} & 2019 & -- & white-box & gradient\\
  \hline
  \cite{suciu2018exploring}& 2018 & -- & white-box & gradient\\
  \hline
  \cite{demetrio2020adversarial}& 2020 & RAMEN & white-box & gradient\\
  \hline
  \cite{hu2017malgan} & 2017 & MalGAN & black-box & GAN\\
  \hline
  \cite{kawai2019improved-malgan} & 2019 & Improved-MalGAN & black-box & GAN\\
  \hline
  \cite{yuan2020gapgan} & 2020 & GAPGAN & black-box & GAN\\
  \hline
\end{tabular}
\end{table*}

%------------------------------------------------------------------
\subsection{Evolutionary Algorithm-based Attacks}
In \cite{castro2019aimed}, an AIMED system was designed and implemented to generate adversarial examples using genetic programming by Castro et al. The system enables the automatic finding of optimized modifications that are applied to previously detected malware and lead to its incorrect evaluation by the malware classifier. It is ensured that all generated adversarial examples are valid PE files. The system implements genetic operations such as selection, crossover, and mutation. If modified PE malware cannot bypass the malware detector, genetic operations are repeated until the generated adversary PE malware can bypass the malware classifier. Experiments have shown that the time to generate successful adversarial examples is reduced by up to 50\% compared to random approaches. 
Furthermore, adversarial examples generated using the given malware classifier were shown to be successful against other malware detectors in 82\% of cases.

Wang and Miikkulainen proposed an adversarial malware detection model named MDEA \cite{wang2020mdea}. This model combines the convolutional neural network to classify raw data from malicious binary and evolutionary optimization to modify detected malware. The action space consists of 10 different methods to modify binary programs. The genetic algorithm evolves different action sequences by selecting actions from the action space until the generated adversarial malware can bypass the target malware detectors. After the successful discovery of action sequences, these sequences are applied to the corresponding malware samples and create a new training set for the detection model. Unlike AIMED, malware samples generated by MDEA are not tested for functionality.

Demetrio et al. introduced a black-box attack framework called GAMMA \cite{demetrio2021gamma}. The black-box attack is the most challenging case since the attacker knows nothing about the target classifier besides the final prediction label. GAMMA attacks are based on the principle of injecting harmless content extracted from goodware into malicious files. Harmless content is inserted into some newly created section or at the end of the file, while the functionality of the file is preserved. The attack is formalized as a constrained optimization problem that minimizes the probability of escaping detection and also penalizes the size of the injected content through a specific penalty.

%------------------------------------------------------------------
\subsection{Reinforcement Learning-based Attacks}
Anderson et al. focused on automating the manipulation of malicious PE files in \cite{anderson2018gym-malware}. The goal is to modify the original malicious PE file so that it is no longer detected as malicious and, at the same time, its format and functionality are not violated. They proposed an attack known as Gym-malware. This is a black-box attack based on reinforcement learning. The authors defined the RL agent's action space as a set of binary manipulation actions. Over time, the agent learns which combinations of actions make malware more likely to bypass antivirus systems.

Song et al. proposed a MAB-Malware framework \cite{song2020mab-malware} based on reinforcement learning to generate adversarial PE malware examples. The action selection problem is modeled as a multi-armed bandit problem. The results showed that the MAB-Malware framework achieves an evasion rate of 74\% to 97\% against machine learning detectors (EMBER \cite{anderson2018EMBER} and MalConv \cite{raff2017malware}) and an evasion rate of 32\% to 48\% against commercial antivirus (AV). Furthermore, they also showed that the transferability of adversarial attacks between ML-based classifiers (i.e., adversarial examples generated against one classifier can be used successfully against another) is greater than 80\%, and the transferability of attacks between pure ML and commercial AV is only up to 7\%.

Fang et al. proposed a framework named DQEAF \cite{fang2019dqeaf} that uses reinforcement learning to evade antimalware engines. DQEAF is similar in methodology to Gym-malware but has many benefits and a higher evasion success rate of adversarial examples compared to it. DQEAF uses a subset of modifications used in Gym-malware and ensures that all modifications will not cause corruption in the modified malware files. DQEAF is able to reduce instability caused by higher dimensions by representing executable files with only 513 features, which is much lower than that in Gym-malware. DQEAF takes priority into account when replaying past transitions. This helps to replay higher-value transitions more frequently, and thus optimize RL networks.

Another reinforcement learning approach is presented in \cite{labaca2021aimed-rl}, in which Labaca-Castro et al. presented the AIMED-RL adversarial attack framework. This attack can generate adversarial examples that lead machine learning models to misclassify malicious files without compromising their functionality. The authors demonstrated the importance of a penalty technique and introduced a new penalization for the reward function with the aim of increasing the diversity of the generated sequences of modifications while minimizing the number of modifications. The results showed that the agents with penalty outperform the agents without penalty in terms of both the best and the average evasion rates.

%------------------------------------------------------------------
\subsection{Gradient-based Attacks}

Kolosnjaji et al. in \cite{kolosnjaji2018amb} introduced a gradient-based white-box attack to generate adversarial malware binaries against MalConv. A white-box scenario occurs when the attacker gets access to the system and may examine its internal configuration or training datasets. The basic idea of the attack is to manipulate some bytes of each malicious software to maximize the likelihood that the input samples are classified as benign. To ensure that malicious binary functionality is maintained, only padding bytes at the end of the file are considered. The results show that the evasion rate is linearly correlated with the length of the injected sequence and, despite the fact that less than 1\% of the bytes are modified, the modified binary evades the target network with high probability and that the precision of MalConv can be reduced by more than 50\%.

In \cite{kreuk2018deceiving} Kreuk et al. presented an improved gradient-based white-box attack method against MalConv. The authors proposed two methods for inserting a sequence of bytes into a file; the payload is inserted either at the end of the file or into slack regions. Unlike \cite{kolosnjaji2018amb}, the evasion rate of \cite{kreuk2018deceiving} is invariant to the length of the injected sequence.

Demetrio et al. in \cite{demetrio2019explaining} presented a gradient-based variant white-box attack that is similar to \cite{kolosnjaji2018amb}. The main difference is that \cite{kolosnjaji2018amb} injects adversarial bytes to the end of the PE file, while this attack is limited to changing bytes within a specific disk operating system (DOS) header in the PE header. The results show that a change of a few bytes is sufficient to evade MalConv with high probability.

Suciu et al. in \cite{suciu2018exploring} describe the FGM append and the FGM slack attacks and compare their effectiveness against MalConv. Experimental results show that the FGM slack attack achieves better results than the FGM append attack with fewer modified bytes.

Demetrio et al. in \cite{demetrio2020adversarial} propose RAMEN, a general adversarial attack framework against PE malware detectors. This framework generalizes and includes previous attacks against machine learning models, as well as three new attacks based on manipulations of the PE file format that preserve its functionality. The first attack is a full DOS attack, which edits all the available bytes in the DOS header. The second attack is called Extend, which enlarges the DOS header, thus enabling manipulation of the extra DOS bytes. The third is the Shift attack, which shifts the content of the first section, creating additional space for the adversarial payload.

%------------------------------------------------------------------
\subsection{Generative Adversarial Network-based Attacks}
Hu and Tan proposed in \cite{hu2017malgan} a model called MalGAN, which is based on a generative adversarial network (GAN). This model enables the generation of adversarial malware examples that are capable of bypassing black-box ML-based detection models. MalGAN includes two neural networks: a generator and a substitute detector. The generator is used to generate adversarial examples, which are dynamically generated according to the feedback of the malware detector and are able to fool the substitute detector. The results showed that MalGAN can reduce the accuracy of detector to almost zero.

Later, Kawai et al. presented Improved MalGAN in \cite{kawai2019improved-malgan}. The authors discuss the problems of MalGAN and try to improve them. For example, the original MalGAN uses multiple malware samples to train MalGAN, in contrast, the improved MalGAN uses only one malware sample. Furthermore, the original MalGAN trains the generator and the malware detector with the same application programming interface (API) call list, while the Improved MalGAN trains with different API call lists.

Yuan et al. in \cite{yuan2020gapgan} introduced GAPGAN, a GAN-based black-box adversarial attack framework. GAPGAN allows end-to-end black-box attacks at the byte level against deep learning-based malware binaries detection. In this approach, a generator and a discriminator are trained concurrently. The generator is used to generate adversarial payloads that are appended to the end of the original data to craft a malware adversarial sample while ensuring the preservation of their functionality. The discriminator tries to imitate the black-box malware detector to recognize both the original benign samples and the adversarial samples generated. When the training process is completed, the trained generator is able to generate an adversarial sample in less than 20 milliseconds. Experiments show that GAPGAN is capable of achieving a 100\% attack success rate against the MalConv malware detector by only inserting adversarial payloads with the size of 2.5\% of the total length of the input malware samples.

%--- EXPERIMENTS --------------------------------------------------
\section{Experiments} \label{sec:experiments}

This section describes the setup and procedure for each experiment. First, the hardware configuration is introduced, followed by a description of the datasets used. Next, the setup of the different algorithms used to generate adversarial samples is described. Finally, the experiments performed are described.

\subsection{Experimental Setup}

The experiments are carried out on the NVIDIA DGX Station A100 server. This server is equipped with an AMD EPYC 7742 processor with a base frequency of 2.25 GHz, 64 cores and 512 GiB of RAM. We also use a virtual machine with Windows 11 operating system and another virtual machine with Kali Linux operating system for testing and analysis.

We use two datasets for our experiments. The first dataset contains a total of 3,625 harmless executable files, which were collected from a newly installed Windows 11 system. The second dataset contains 3,625 malicious executable files obtained from the VirusShare repository \cite{virus-share}. All the files in both datasets are PE files.

\subsection{Attack settings}
In total, we compare five adversarial attack strategies. Partial DOS and Full DOS attacks are performed in a white-box setting against the MalConv detector and the maximum number of iterations was set to 50. GAMMA padding and GAMMA section-injection attacks are performed in a black-box setting against the MalConv detector with the maximum number of queries set to 500, regularization parameter set to $10^{-5}$, and a total of 100 .data sections extracted from benign programs used as injection content. The last attack we use is the Gym-malware attack with its default settings performed in a black-box setting against the GDBT detector. The Gym-malware model was trained on a dataset of 3,000 malicious samples and a validation set of 1,000 files.

\subsection{Evaluation Metrics} \label{evaluation-metrics}

We use several metrics to evaluate the experiments. A key metric in the area of adversarial machine learning is the evasion rate ($ER$), the proportion of malware files misclassified by the target malware classifier, and can be calculated as follows:
\begin{equation}
ER = \frac{misclassified}{total}
\end{equation}
where $misclassified$ is the number of malware samples misclassified as benign and $total$ is the total number of files submitted to the target classifier after discarding files that were already incorrectly predicted before modification.

The evasion rate mentioned above is a universal metric that can be used to evaluate both single attacks and combinations of attacks. In both cases, we are interested in the percentage of malware that escaped detection by the antivirus program. Additionally, we use the following metrics to evaluate the combination of attacks.

The first two metrics that we chose to evaluate the success of the combination are the absolute improvement and the relative improvement in the evasion rate when using the second attack in the combination compared to the first attack. Absolute improvement ($AI$) can be described by the following formula:
\begin{equation} \label{eq:absolute-improvement}
AI = ER_C - ER_1
\end{equation}
where $ER_C$ is the total evasion rate when using a combination of methods, and $ER_1$ is the evasion rate after using the first attack in the combination alone. The result is the percentage increase in evasion rate between the first and second attack in the combination. For example, if the evasion rate after the first attack in the combination is $0.01$ and after the second attack is $0.1$, then the absolute improvement is $0.09$, which means that the second attack improved the overall evasion rate by $9\%$.  

Similarly, the relative improvement ($RI$) can be expressed using the formula: 
\begin{equation}
RI = \frac{ER_C - ER_1}{ER_1}
\end{equation}
where the meaning of the variables is the same as in the previous formula \eqref{eq:absolute-improvement}. However, in the previous case, the result expressed a percentage increase over all samples tested. In the case of relative improvement, we limit ourselves to the set of samples that escaped the antivirus program after the first attack. For example, if the evasion rate after the first attack of the combination is $0.01$ and after the second attack it is $0.1$, then the relative improvement is $9$, i.e. the second attack improved the evasion rate of the first attack by $900\%$.  

Next, we need to compare the combination of attacks with performing the attacks separately to see if the combination of attacks adds any value. To do this, we use a simple comparison of the evasion rate of the combination of attacks with the evasion rate of the attacks performed separately. We call it evasion rate comparison ($ERC$), and it has the following calculation:

%To do this, we use two metrics, the first of which we will call the evasion rate benefit ($ERB$), and it is defined as follows:
%\begin{multline}
%ERB = \\\frac{mis_C - mis_{C,1} - mis_{C,2} + mis_{C,1,2}}{total}
%\end{multline}
%where $mis_C$ is the number of malware not detected as malware after the combination of attacks, $mis_{C,1}$ represents the number of malware that were not detected as malware after executing both the attack combination and the first attack alone, similarly $mis_{C,2}$ for the second attack. Furthermore, $mis_{C,1,2}$ is the number of malware counted under both $mis_{C,1}$ and $mis_{C,2}$. The denominator $total$ is the number of all malware in the test set. Thus, the result generally reflects the percentage of samples that escaped detection due to a combination of attacks; separate attacks performed on these samples would have failed. If the result is 0, then the combination did not add value over the first and second attacks performed separately.

%The final metric is a simple comparison of the evasion rate of the combination of attacks with the evasion rate of the attacks performed separately. We call it evasion rate comparison ($ERC$), and it has the following calculation:
\begin{equation}
ERC = ER_C - \max( ER_1, ER_2 )
\end{equation}
where $ER_C$ is the evasion rate of the combination attack, while $ER_1$ and $ER_2$ are the evasion rates of the first and second attacks, respectively. If the result is positive, it means that the combination of attacks performed better than the combination of the two attacks in the combination that would have been performed alone. If the result is negative or zero, it means that the execution of the attack combination was pointless because one of the attacks that were part of the combination performed better or was equal to the combination in terms of evasion rate.

\subsection{Experiments Description}
We present four experiments that explore different characteristics of the adversarial attack methods mentioned above.

\subsubsection{Sample Generation Time}
In the first experiment, we measure the time it took to generate individual samples using all the aforementioned selected algorithms. These results, along with other data collected during the experiments, can help compare the effectiveness of various generators and determine the most effective method to generate adversarial malware samples.

\subsubsection{Sample Size}
In the second experiment, we investigate how the size of the original malware samples changes by applying various adversarial malware generators. Generally, the attacker's goal is to minimize the increase in the size of the generated adversarial files to make it harder to distinguish them from the original malware samples.

\subsubsection{Bypassing commercial AV products}
In the third experiment, we analyze the effectiveness of created adversarial malware samples against real-world AV detectors. Based on a comparative study \cite{av-comparatives}, 10 AV programs were selected for experimentation, and their names were intentionally anonymized in the following results. Note that in the subsequent results, only nine AVs are listed as two of the selected AVs reported the same results.

The modified malware files from different adversarial algorithms are submitted to the VirusTotal server \cite{virus-total} to obtain the evasion rate for each adversarial malware generator. To avoid bias in the results, we only analyzed malware samples that were correctly classified by all selected AV products before modification. We also discarded samples from which we were unable to obtain file analysis from VirusTotal, e.g., due to broken behavior of modified examples. In total, we use a set of 530 genuine malware samples along with a modified version for each malware generator.

\subsubsection{Combination of Multiple Techniques} \label{sec444}
In the last experiment, we test the effectiveness of using a combination of methods to generate malware samples \cite{kozak2023combining}. The goal of this experiment is to test whether using multiple adversarial example generators per malware sample would significantly increase the malware evasion rate.

\begin{figure}[h]
    \centering
    \includegraphics[width=0.35\textwidth]{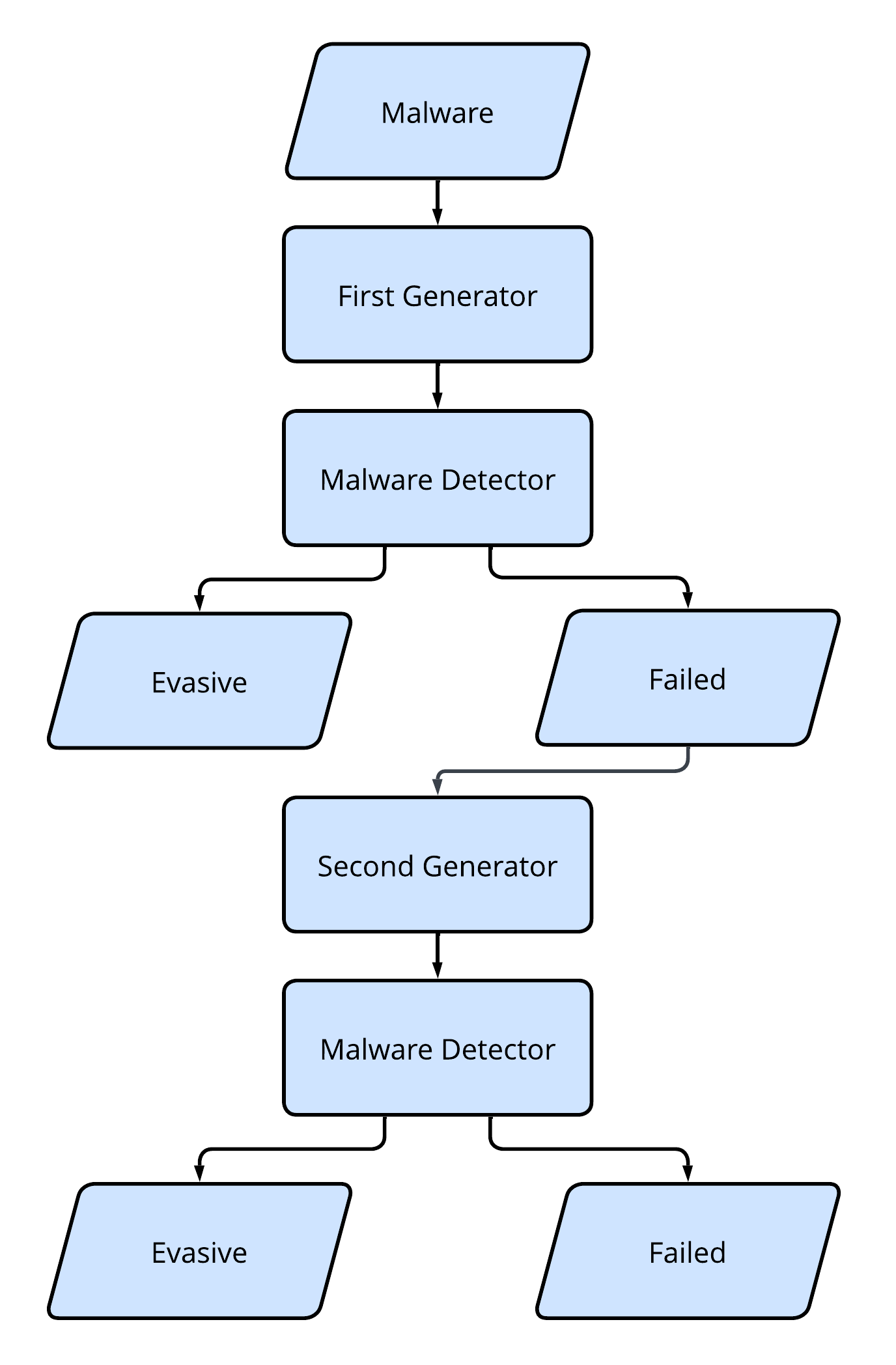}
    \caption{Method for generating adversarial examples by combining two generators.}
    \label{fig:mix}
\end{figure}

An overview of the experiment is shown in Figure \ref{fig:mix}. First, the original malware samples are processed by the first generator. These modified samples are then tested against a real AV detector that is not part of the generator. The result is a set of samples divided into two sets. The first set consists of \textit{evasive examples} that successfully evaded the given malware detector, and this set is no longer processed. In contrast to \textit{adversarial examples}, which are generated against the target classifier, \textit{evasive examples} are samples that have evaded detection by the AV program, although this AV program was not used to generate these samples. The second set consists of \textit{failed examples} that failed to evade the detector and are used as input to the second generator. The second generator processes the \textit{failed examples}, and the resulting modified samples are again tested against real the AV detector. The result is again a set of samples divided into two sets: \textit{evasive} and \textit{failed}. The set of \textit{evasive examples} produced by the first generator and the set of \textit{evasive examples} produced by the second generator together form the set of resulting successful adversarial examples produced by combining these two generators. The \textit{failed examples} obtained after using the second generator are the resulting samples that did not evade detection.

%--- RESULTS -----------------------------------------------------
\section{Results} \label{sec5}
This section presents the recorded results of the individual experiments described in Section \ref{sec:experiments}.

\subsection{Sample generation time}
\begin{figure}[h]
    \includegraphics[width=0.5\textwidth]{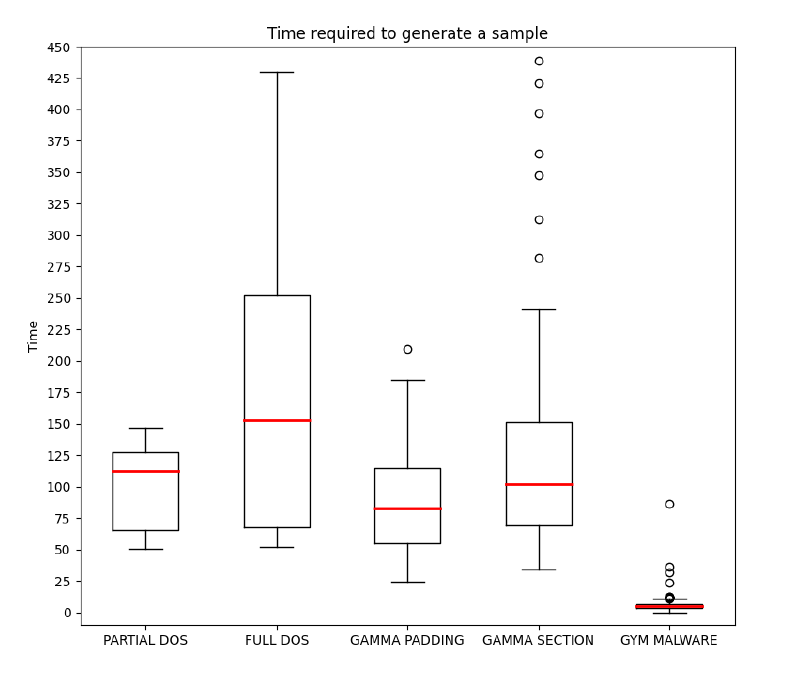}
    \caption{Time required to generate a sample for each sample generator.}
    \label{fig:time}
\end{figure}

Firstly, we look at the results of the sample generation time. Average times in seconds and standard deviation for each attack are listed in Table \ref{tab:time} and in the box plot in Figure \ref{fig:time}.

\begin{table*}[h]
\centering
\caption{Average sample generation time for each sample generator.}
\label{tab:time}
%\resizebox{0.47\textwidth}{!}{%
\begin{tabular}{|l|r|r|}
\hline
Attack                  & Average Duration [s] & Standard Deviation [s]\\ \hline \hline
Partial DOS             & $99.27$        & $31.13$  \\ \hline
Full DOS                & $169.08$       & $104.53$ \\ \hline
GAMMA Padding           & $87.61$        & $39.28$  \\ \hline
GAMMA Section-injection & $118.47$       & $69.44$  \\ \hline
Gym-malware             & $5.73$         & $7.52$   \\ \hline
\end{tabular}%
%}
\end{table*}

From Figure \ref{fig:time}, we can see that Gym-malware took the least amount of time, on average less than 6 seconds, to generate adversarial examples with some outliers below 100 seconds. It should be noted that Gym-malware requires a preceding training phase that is not taken into account in this experiment. On the contrary, the Full DOS attack recorded the longest time to create adversarial examples, with an average duration of more than 160 seconds. The remaining three methods achieved similar sample generation times of around 100 seconds in spite of the fact that the results of the GAMMA section-injection attack contain several outliers with extensive long durations of more than 300 seconds. However, the measured times may be affected by the settings of individual algorithms.

\subsection{Sample Size}
Secondly, we present the results of the sample size experiment. Partial DOS and Full DOS attacks are based on changing bytes in the DOS header. Thus, modifying the malware samples by these methods does not alter the size of the resulting adversarial examples. On the other hand, the remaining tested methods can change the initial file size. The general results of this experiment are shown in Table \ref{tab:size}.

The GAMMA padding and section-injection attacks are based on inserting parts extracted from benign files into the malware file. In the case of the GAMMA padding attack, the file size increased on average by 223,605 bytes and in the case of the GAMMA section-injection attack, by 1,940,352 bytes. As we can see from Table \ref{tab:size}, the adversarial examples generated by the GAMMA section-injection attack exhibit significant differences in the final file sizes. 

The Gym-malware attack uses various types of file manipulations. The file size can be reduced, increased, or unchanged based on the chosen modification. On average, the file size was reduced by 149,273 bytes. The observed file size reduction was probably due to the authors' implementation of file manipulations, as they used the LIEF library \cite{LIEF}, which significantly alters the structure of the initial malware file.

\begin{table*}[h]
\centering
\caption{Changes to the sample size of generated adversarial samples for each generator.}
\label{tab:size}
%\resizebox{0.47\textwidth}{!}{%
\begin{tabular}{|l|r|r|}
\hline
Attack                  & Average Size [B]  & Standard Deviation [B]\\ \hline \hline
Partial DOS             & $0$           & $0$                       \\ \hline
Full DOS                & $0$           & $0$                       \\ \hline
GAMMA Padding           & $223,604.57$   & $48,403.25$              \\ \hline
GAMMA Section-injection & $1,940,351.63$  & $78,088,981.28$         \\ \hline
Gym-malware             & $-149,272.75$  & $754,660.5$              \\ \hline
\end{tabular}%
%}
\end{table*}

\subsection{Bypassing commercial AV products}
\begin{table*}[ht!]
  \centering
  \caption{Evasion rate (in \%) of generated adversarial samples achieved against AV products on the VirusTotal server.}
  \label{tab:av}
  \resizebox{1\textwidth}{!}{%
  \begin{tabular}{|l||l|l|l|l|l|l|l|l|l||l|}
  \hline
  Attack & AV1  & AV2 & AV3  & AV4  & AV5  & AV6  & AV7  & AV8  & AV9  & Average \\ \hline \hline
  GAMMA padding & 0.00 & 1.79 & 0.45 & 0.22 & 0.45 & 0.90 & 1.34 & 0.56 & 0.45 & 0.68 \\ \hline
  Partial DOS & 0.78 & 2.57 & 0.78 & 1.01 & 0.78 & 0.78 & 1.90 & 1.45 & 0.78 & 1.21 \\ \hline
  Full DOS & 0.67 & 1.34 & 0.78 & 0.90 & 0.78 & 0.78 & 4.14 & 1.23 & 0.78 & 1.27 \\ \hline
  GAMMA section-injection & 18.46 & 5.37 & 6.38 & 4.36 & 4.47 & 9.06 & 43.62 & 1.23 & 5.37 & 10.92 \\ \hline
  Gym-malware & 45.53 & 19.02 & 44.86 & 67.23 & 41.61 & 53.58 & 53.80 & 26.51 & 44.86 & 44.11 \\ \hline 
  \end{tabular}%
  }
\end{table*}

Thirdly, we list the effectiveness of generated adversarial samples on commercially available AV programs. The results are shown in Table \ref{tab:av}. Each column represents one of the selected AV programs and each row represents one of the algorithms used to generate adversarial samples. The values in the table represent the achieved evasion rates, expressed as a percentage, for the corresponding algorithm and AV products.

The Gym-malware attack achieved the highest evasion rates among all selected AV programs, successfully bypassing the top AVs 19.02\% to 67.23\% of the time. The second-best results were recorded by the GAMMA section-injection attack, which recorded evasion rates between 1.23\% and 43.62\%.

In contrast, the GAMMA padding attack achieved the worst results, failing to mislead any detector tested in more than 1.5\% of cases. Full and Partial DOS attacks scored slightly better than the GAMMA padding attack, with the Full DOS marginally outperforming the Partial DOS attack.

\subsection{Combination of Multiple Techniques}
Based on the results from the previous experiment, we chose the three most successful adversarial example generators and tested all nine possible combinations. Namely, we used Gym-malware, GAMMA section-injection, and Full DOS adversarial malware generators.

This section contains two types of tables. The first type of table lists the measured minimum, average, and maximum values for particular AVs across the nine combinations of the three selected generators. The second type of table lists the measured minimum, average, and maximum values for a particular combination of generators across the nine AVs. The First Generator column contains the first generator used, and the Second Generator column contains the second generator used as described in Section \ref{sec444}.

\subsubsection*{Evasion Rate}

\begin{table}[h]
  \centering
  \caption{Evasion rate (in \%) for each AV using all combinations of generators.}
  \label{evasion-rate-avs}
  %\resizebox{0.25\textwidth}{!}{%
  \begin{tabular}{|l||r|r|r|}
  \hline
  AV  & Minimum  & Average  & Maximum   \\ \hline \hline
  AV1 & 0.78 & 32.39 & 55.26 \\ \hline
  AV2 & 1.45 & 17.79 & 29.53 \\ \hline
  AV3 & 0.90 & 31.15 & 63.09 \\ \hline
  AV4 & 1.45 & 38.59 & 78.19 \\ \hline
  AV5 & 0.78 & 26.76 & 57.61 \\ \hline
  AV6 & 0.90 & 35.91 & 73.60 \\ \hline
  AV7 & 5.26 & 49.32 & 74.50 \\ \hline
  AV8 & 1.57 & 17.80 & 41.39 \\ \hline
  AV9 & 0.78 & 30.79 & 62.75 \\ \hline
  \end{tabular}%
  %}
\end{table}

\begin{table*}[h!]
  \centering
  \caption{Evasion rate (in \%) for each generator combination against all AVs.}
  \label{evasion-rate-generators}
  %\resizebox{0.45\textwidth}{!}{%
  \begin{tabular}{|l|l||r|r|r|}
  \hline
  First Generator         & Second Generator        & Minimum   & Average  & Maximum   \\ \hline \hline
  Full DOS                & Full DOS                & 0.78  & 1.54  & 5.26  \\ \hline
  Full DOS                & GAMMA section-injection & 1.57  & 10.48 & 43.96 \\ \hline
  Full DOS                & Gym-malware             & 23.15 & 38.69 & 61.63 \\ \hline
  GAMMA section-injection & Full DOS                & 6.26  & 15.05 & 45.30 \\ \hline
  GAMMA section-injection & GAMMA section-injection & 1.90  & 14.03 & 44.52 \\ \hline
  GAMMA section-injection & Gym-malware             & 25.39 & 46.97 & 74.50 \\ \hline
  Gym-malware             & Full DOS                & 26.51 & 46.16 & 67.34 \\ \hline
  Gym-malware             & GAMMA section-injection & 27.18 & 49.22 & 67.79 \\ \hline
  Gym-malware             & Gym-malware             & 29.53 & 58.34 & 78.19 \\ \hline
  \end{tabular}%
  %}
\end{table*}

First, we present the results of the evasion rate metric. For all AVs, we examine the minimum, average, and maximum of the results of all combinations of generators tested in the experiment. These results can be found in Table \ref{evasion-rate-avs}. For the minimum values of the evasion rate, we can see that none of the antivirus programs reached a detection rate of 100\%. On the other hand, all these values are relatively low compared to the average and maximum values, which tells us that some of the nine generator combinations were not very successful. The average evasion rates range from about 18\% to 49\%, and the maximum values range from 30\% to 78\%. The best result was achieved by combinations of generators against AV7, where the average evasion rate is around 49\%, while the least successful was against AV2, where the average evasion rate is around 18\%.
%\todo{This means that if we use all combinations in the experiment, we achieve an average evasion rate of at least 18\% for all selected AVs, and some of the combinations achieve an evasion rate of around 30\% for all AVs.}

%Table \ref{evasion-rate-avs} shows the results of the evasion rate for each AV. Table \ref{evasion-rate-generators} shows the results for each generator combination. Both tables contain aggregated data, specifically the minimum, average, and maximum values across all AVs and all generator combinations, respectively.

Table \ref{evasion-rate-generators} shows the results of the evasion rate achieved by each combination of generators. Here, we can see that the most successful combination was the one in which the Gym-malware generator was used twice in a row. This achieved a minimum evasion rate of around 30\% for all AVs. The average value for this combination is around 58\%, and for at least one AV we achieved an evasion rate of around 78\% with this combination. On the other hand, the worst combination in terms of evasion rate is the one in which the Full DOS generator was used twice. In this case, we have an average evasion rate of about 2\%, while for all other combinations, this value exceeds 10\%, and for some even more significantly. The maximum value of the evasion rate for this combination is about 5\%, for the others we have at least about 44\%.

To determine the effectiveness of using generator combinations instead of individual generators alone, we can compare these results with the values from the previous experiment. However, we have additional metrics for this, which are described in Section \ref{evaluation-metrics}. We analyze the results of these metrics in the following parts of this section.

\subsubsection*{Absolute Improvement}

Next, we evaluate the metric that we identified as an absolute improvement in Section \ref{evaluation-metrics}. In short, it is the percentage difference in the evasion rate between the evasion rate achieved by combining both generators and the evasion rate achieved by using only the first generator.
\begin{table}[h]
  \centering
  \caption{Absolute improvement (in \%) for each AV using all combinations of generators.}
  \label{absolute-improvement-avs}
  %\resizebox{0.25\textwidth}{!}{%
  \begin{tabular}{|l||r|r|r|}
  \hline
  AV  & Minimum  & Average  & Maximum   \\ \hline \hline
  AV1 & 0.11 & 10.84 & 29.98 \\ \hline
  AV2 & 0.11 & 9.21  & 22.48 \\ \hline
  AV3 & 0.11 & 13.81 & 41.16 \\ \hline
  AV4 & 0.00 & 14.43 & 60.74 \\ \hline
  AV5 & 0.00 & 11.14 & 35.01 \\ \hline
  AV6 & 0.11 & 14.77 & 50.22 \\ \hline
  AV7 & 0.90 & 15.46 & 40.72 \\ \hline
  AV8 & 0.34 & 8.14  & 26.51 \\ \hline
  AV9 & 0.00 & 13.78 & 41.39 \\ \hline
  \end{tabular}%
  %}
\end{table}

\begin{table*}[h!]
  \centering
  \caption{Absolute improvement (in \%) for each generator combination against all AVs.}
  \label{absolute-improvement-generators}
  %\resizebox{0.45\textwidth}{!}{%
  \begin{tabular}{|l|l||r|r|r|}
  \hline
  First Generator         & Second Generator        & Minimum   & Average  & Maximum   \\ \hline \hline
  Full DOS                & Full DOS                & 0.00  & 0.27  & 1.12  \\ \hline
  Full DOS                & GAMMA section-injection & 0.67  & 9.21  & 39.82 \\ \hline
  Full DOS                & Gym-malware             & 21.92 & 37.42 & 60.74 \\ \hline
  GAMMA section-injection & Full DOS                & 0.78  & 4.13  & 5.93  \\ \hline
  GAMMA section-injection & GAMMA section-injection & 0.00  & 3.11  & 10.29 \\ \hline
  GAMMA section-injection & Gym-malware             & 20.02 & 36.04 & 51.23 \\ \hline
  Gym-malware             & Full DOS                & 0.11  & 2.05  & 7.49  \\ \hline
  Gym-malware             & GAMMA section-injection & 0.56  & 5.11  & 10.63 \\ \hline
  Gym-malware             & Gym-malware             & 7.72  & 14.23 & 20.02 \\ \hline
  \end{tabular}%
  %}
\end{table*}

First, we focus on Table \ref{absolute-improvement-avs}, which shows the absolute improvement values for each AV in the nine combinations of generators. Here we can see that for some AVs we were unable to improve the evasion rate by applying the second generator. Specifically, this refers to the minimum value for AV4, AV5, and AV9. Table \ref{absolute-improvement-generators} helps us to identify the relevant generators. We can see that the only non-improving combinations are the ones in which we use the same generator twice, namely, the Full DOS generator or the GAMMA section-injection generator. This may indicate that the use of these combinations is not entirely effective. For the average values in Table \ref{absolute-improvement-avs}, we can see that they do not differ significantly, in all cases between about 8\% and 15\%. For the maximum values, we have a slightly higher range, approximately 22\% to 61\%. This means that if we use a second generator, we will improve the evasion rate by about 10\% on average after using the first generator. %\todo{If we use all combinations, we will achieve an improvement in the evasion rate of at least 22\% for all AVs.}

However, significantly more intriguing is Table \ref{absolute-improvement-generators}, which shows the absolute improvement for each generator combination against all AVs. Here, we can see that the use of Full DOS as the second generator does not result in a significant absolute improvement in the evasion rate for the minimum, average, and maximum values. This means that using Full DOS as the second generator in a combination is the least effective. On the other hand, we can see that we get the best absolute improvement by using the Gym-malware method as the second generator in the combination. %The GAMMA section-injection as the second generator does not have the best or worst results. We could also see this in the previous section regarding the evasion rate, where these combinations did not achieve the best result in any case.

In Table \ref{absolute-improvement-generators}, we can observe another interesting fact. As we have already mentioned, the use of two identical generators in combination is not very effective. However, this statement does not apply to the Gym-malware generator. In contrast, if we use the Gym-malware generator as the first generator in the combination, then the best choice to select the second generator seems to be the Gym-malware generator again. Full DOS and GAMMA section-injection generators have minimal absolute improvements in the role of the second generator. These results show that the Gym-malware generator is successful in both cases, used alone and in combination with itself. %even when used alone, combining generators does not improve it much, in the case that this generator was used as the first generator in the combination.

The best absolute improvement values are achieved when we choose Full DOS as the first generator and Gym-malware as the second. This results in an absolute improvement in the evasion rate in all AVs of at least around 22\%, on average 37\% and up to a maximum of 61\%. On the other hand, the worst results in terms of absolute evasion rate improvement are obtained when we choose Full DOS as both generators. In this case, we obtain a minimum absolute improvement of 0\%, an average of 0.3\% and a maximum of 1.1\% across all antivirus programs. We can conclude that the Full DOS generator is likely to be the least successful generator, both when used alone and in combination.

\subsubsection*{Relative Improvement}

\begin{table}[h]
  \centering
  \caption{Relative improvement (in \%) for each AV using all combinations of generators.}
  \label{relative-improvement-avs}
  %\resizebox{0.25\textwidth}{!}{%
  \begin{tabular}{|l||r|r|r|}
  \hline
  AV  & Minimum  & Average  & Maximum   \\ \hline \hline
  AV1 & 0.61 & 788.88 & 4466.67 \\ \hline
  AV2 & 8.33 & 307.66 & 1675.00 \\ \hline
  AV3 & 3.74 & 732.66 & 5014.29 \\ \hline
  AV4 & 0.00 & 914.67 & 6787.50 \\ \hline
  AV5 & 0.00 & 633.29 & 4257.14 \\ \hline
  AV6 & 1.46 & 846.29 & 6000.00 \\ \hline
  AV7 & 2.05 & 232.70 & 983.78  \\ \hline
  AV8 & 3.80 & 510.39 & 2154.55 \\ \hline
  AV9 & 0.00 & 758.01 & 5285.71 \\ \hline
  \end{tabular}%
  %}
\end{table}

\begin{table*}[h]
  \centering
  \caption{Relative improvement (in \%) for each generator combination against all AVs.}
  \label{relative-improvement-generators}
  %\resizebox{0.45\textwidth}{!}{%
  \begin{tabular}{|l|l||r|r|r|}
  \hline
  First Generator         & Second Generator        & Minimum    & Average    & Maximum     \\ \hline \hline
  Full DOS                & Full DOS                & 0.00   & 18.93   & 62.50   \\ \hline
  Full DOS                & GAMMA section-injection & 75.00  & 715.51  & 2400.00 \\ \hline
  Full DOS                & Gym-malware             & 983.78 & 4027.99 & 6787.50 \\ \hline
  GAMMA section-injection & Full DOS                & 3.85   & 104.90  & 409.09  \\ \hline
  GAMMA section-injection & GAMMA section-injection & 0.00   & 58.50   & 181.25  \\ \hline
  GAMMA section-injection & Gym-malware             & 70.77  & 741.93  & 2154.55 \\ \hline
  Gym-malware             & Full DOS                & 0.17   & 7.31    & 39.41   \\ \hline
  Gym-malware             & GAMMA section-injection & 0.83   & 13.60   & 42.94   \\ \hline
  Gym-malware             & Gym-malware             & 16.31  & 35.90   & 56.12   \\ \hline
  \end{tabular}%
  %}
\end{table*}

We follow the absolute improvement results with a relative improvement in the evasion rate. Analogously to the improvement in the absolute evasion rate, we can see the results of relative improvement in Table \ref{relative-improvement-avs}. We can see that AV7 performed the best on average in terms of relative improvement of the evasion rate. Conversely, AV4 performed the worst, where in some cases the second generator managed to increase the number of successfully modified samples (which evaded detection) by more than 67 times.

%AVs that we did not improve the evasion rate with the second generator. These are still the same AVs, specifically AV4, AV5, and AV9 and combinations of generators consisting of two identical generators Full DOS or GAMMA section-injection.

Table \ref{relative-improvement-generators} confirms that the Full DOS generator, which was used as the second generator in combination, does not increase the evasion rate significantly. On the other hand, if this generator is chosen as the first generator in the combination, the GAMMA section-injection or Gym-malware generator can increase the number of samples that evade detection by up to 67 times. This shows that the Full DOS generator is not a very strong generator on its own. Regarding the Gym-malware generator, we can see that when it is used as the first generator in a combination, the second generator does not significantly increase the result, even when Gym-malware is used again as the second generator. However, the Gym-malware generator is very successful on its own as it achieves a significantly higher evasion rate than other generators. Therefore, a relative improvement in the evasion rate from 39\% to 56\% can cause a drastic increase in the total evasion rate of the combination (up to 20\%). We can also see that when the Gym-malware generator is used as a second generator in other combinations, there are huge relative improvements over the first generator. Again, we can see that Gym-malware is very effective when used in any combination of generators.

\subsubsection*{Evasion Rate Comparison}
The last metric examined, evasion rate comparison, helps us find the answer to the question of whether it is better to use individual generators separately or to use them in combination. A negative value of this metric indicates that we would achieve a better evasion rate by using the better of the two generators individually (better in terms of evasion rate). On the contrary, a positive value indicates that we recorded a better result using the combination of both generators.

%This metric measures the level of leakage and simply observes whether the result is better when a better generator from the combination is used separately or when a combination of generators is used.
\begin{table}[h]
  \centering
  \caption{Evasion rate comparison (in \%) for each AV using all combinations of generators.}
  \label{evasion-rate-comparison-avs}
  %\resizebox{0.25\textwidth}{!}{%
  \begin{tabular}{|l||r|r|r|}
  \hline
  AV  & Minimum  & Average  & Maximum   \\ \hline \hline
  AV1 & -14.88 & 0.87  & 9.73  \\ \hline
  AV2 & 0.11   & 5.28  & 10.52 \\ \hline
  AV3 & -4.81  & 4.02  & 18.23 \\ \hline
  AV4 & -11.63 & -0.31 & 10.96 \\ \hline
  AV5 & -7.49  & 2.06  & 16.00 \\ \hline
  AV6 & -5.82  & 3.03  & 20.02 \\ \hline
  AV7 & -8.95  & 4.43  & 20.69 \\ \hline
  AV8 & -3.36  & 2.52  & 14.88 \\ \hline
  AV9 & -2.69  & 3.99  & 17.90 \\ \hline
  \end{tabular}%
  %}
  \end{table}
  
\begin{table*}[h]
  \centering
  \caption{Evasion rate comparison (in \%) for each generator combination against all AVs.}
  \label{evasion-rate-comparison-generators}
  %\resizebox{0.45\textwidth}{!}{%
  \begin{tabular}{|l|l||r|r|r|}
  \hline
  First Generator         & Second Generator        & Minimum    & Average  & Maximum   \\ \hline \hline
  Full DOS                & Full DOS                & 0.00   & 0.27  & 1.12  \\ \hline
  Full DOS                & GAMMA section-injection & -2.80  & -0.45 & 1.57  \\ \hline
  Full DOS                & Gym-malware             & -14.88 & -5.42 & 4.81  \\ \hline
  GAMMA section-injection & Full DOS                & 0.78   & 4.13  & 5.93  \\ \hline
  GAMMA section-injection & GAMMA section-injection & 0.00   & 3.11  & 10.29 \\ \hline
  GAMMA section-injection & Gym-malware             & -11.63 & 2.86  & 20.69 \\ \hline
  Gym-malware             & Full DOS                & 0.11   & 2.05  & 7.49  \\ \hline
  Gym-malware             & GAMMA section-injection & 0.56   & 5.11  & 10.63 \\ \hline
  Gym-malware             & Gym-malware             & 7.72   & 14.23 & 20.02 \\ \hline
  \end{tabular}%
  %}
\end{table*}

Looking at the minimum values in table \ref{evasion-rate-comparison-avs}, we can see that for almost all AVs there was a combination of generators that was not effective due to the negative values in this column. The exception is AV2, where we achieved a better evasion rate for all combinations of generators used. However, the average and maximum values show that in most cases the combination is more effective than the better of the two generators used separately. Only AV4 has a negative average value, which means that a separate use of a single generator is a better option against this AV. On the other hand, when attacking AV2 it is better to use a combination of generators as it achieved the highest average evasion rate comparison value.

Table \ref{evasion-rate-comparison-generators} shows that combining generators using Full DOS as the first generator and either GAMMA section-injection or Gym-malware as the second generator yields negative results, as evidenced by both the minimum and average values being negative. Based on the results from the previous parts of this section, we can say that it is more advantageous to use GAMMA section-injection and Gym-malware generators separately in such cases. We can also see other negative values in the minimum value column for the combination of GAMMA section-injection and Gym-malware generators used in this order. Nonetheless, the average value of the combination is positive, indicating that it is beneficial to use this combination on average. For the remaining combinations, we can conclude that using them results in a better evasion rate than using the better of the two generators separately.

%However, further research is needed to confirm this conclusion.  Once again, the combination of two Gym-malware generators was the best in this evaluation.

%--- CONCLUSION --------------------------------------------------
\section{Conclusion} \label{sec6}

In this paper, we explored the use of adversarial learning techniques in malware detection. Our goal was to apply existing methods for generating adversarial malware samples, test their effectiveness against selected malware detectors, and compare the evasion rates achieved and the practical applicability of these methods.

For our experiments, we chose five adversarial malware sample generators: Partial DOS, Full DOS, GAMMA padding, GAMMA section-injection, and Gym-malware. This selection represents a spectrum of adversarial techniques based on gradient, evolutionary algorithms, and reinforcement learning. These adversarial malware generators were evaluated on nine commercially available antivirus products.

To validate and compare the different characteristics and properties of the methods used, we performed four experiments. These included tracking the time taken to generate samples, changes in sample size after using applying adversarial modifications, testing effectiveness against antivirus programs, and evaluating combinations of generators.

The results indicate that making optimized modifications to previously detected malware can cause the classifier to misclassify the file and label it as benign. Furthermore, the study confirmed that generated malware samples could be used successfully against detection models other than those used to generate them. Using combination attacks, a significant percentage of new samples were created that could evade detection by antivirus programs.

%Experiments showed that the Gym-malware generator, which uses a reinforcement learning approach, has the greatest practical potential. This generator produces malware samples in the shortest time, with an average sample generation time of 5.73 seconds. Additionally, the Gym-malware generator achieved the highest evasion rate among all selected antivirus products, with the highest evasion rate recorded at 67\%. Furthermore, the Gym-malware generator was found to be effective when used in combination with another generator, especially with itself, where it achieved the highest evasion rate of up to 78\%.

Experiments showed that the Gym-malware generator, which uses a reinforcement learning approach, has the greatest practical potential. This generator produced malware samples in the shortest time, with an average sample generation time of 5.73 seconds. The Gym-malware generator also achieved the highest evasion rate among all selected antivirus products, with the highest average evasion rate of 44.11\% against nine AVs. Furthermore, the Gym-malware generator was effective when combined with another generator, especially with itself, where it achieved the highest average evasion rate of 58.35\%. Additionally, this generator could significantly improve the performance of other generators with absolute and relative improvements ranging between 36.04\%--37.42\% and 741.93\%--4027.99\%, respectively.

For future experiments, we propose to study in more detail how the sample generation time is affected by the input genuine malware size and investigate the correlation between the time and the evasion rate of the resulting adversarial examples. Further experiments could be done in the area of combining generators where more than two generators could be combined to achieve even higher evasion rates. Our work highlights the importance of developing new techniques to detect malware and identify adversarial attacks. More research is needed in this area to successfully combat these novel threats and attacks.

\section*{\uppercase{Acknowledgements}}
This work was supported by the Student Summer Research Program 2022 of FIT CTU in Prague and by the OP VVV MEYS funded project CZ.02.1.01/0.0/0.0/16 019/0000765 ``Research Center for Informatics'' and by the Grant Agency of the CTU in Prague, grant No. SGS23/211/OHK3/3T/18 funded by the MEYS of the Czech Republic.

\section*{Declarations}

The authors have no relevant financial or non-financial interests to disclose.

\bibliography{sn-article}% common bib file

\end{document}